\documentclass[a4paper,12pt]{article}
\usepackage{epsfig}
\begin{document}
\baselineskip=17pt
\title{Curvature-induced phase transition in three-dimensional 
Thirring model}
\author{B. Geyer\thanks{e-mail: geyer@ntz.uni-leipzig.d400.de}\\
{\it Institute of Theoretical Physics and Center for Advanced Studies,}\\
{\it Leipzig University, Augustusplatz 10, D--04109 Leipzig, 
Germany}\\[1.5ex]
Yu. I. Shil'nov\thanks{e-mail: shilnov@itp.uni-leipzig.de. Permanent e-mail: 
shilnov@express.kharkov.ua.  }\\
{\it Department of Theoretical Physics, Faculty of Physics,}\\
{\it Kharkov State University,}\\
{\it Svobody Sq. 4, 310077, Kharkov, Ukraine}\\
{\it and}\\
{\it Center for Advanced Studies, Leipzig University,}\\
{\it Augustusplatz 10, D--04109, Leipzig, Germany}}
\date{ }
\maketitle
\large
\vskip 1cm

\noindent 
The effective potential of composite fermion fields in three dimensional \linebreak
Thirring model in curved space--time is calculated in linear 
curvature approximation. The phase transition accompanied by the creation of 
non--zero chiral invariant bifermionic vector--like condensate is shown to exist.
The type of this phase transition is discussed.

\vfill\eject

{\bf 1}. As is well established the four-fermion models provide a useful tool for the
description of  low and intermediate energy physics of strong interactions [1-6].
They can be obtained from the full QCD by means of the integration over the 
gluon degrees of freedom. However, it requires the introduction of a 
dimensional scale, limiting the
low energy region where the non--local character of fermion interactions due 
to  
exchange of gluons can be neglected.

Furthermore, the Nambu--Jona-Lasinio (NJL) model has recently been used to introduce a
dynamical symmetry breaking mechanism which is very important for high energy physics 
(see, for example [7, 8] and literature, cited there). 

The most popular and the simplest four-fermion model is the NJL model with the scalar 
and pseudoscalar types of fermion interaction only [1]. It has been frequently applied 
to obtain the dynamical symmetry
breaking and to create bifermionic bound states (mesons) in a manner  appropriate 
for the phenomenology of hadrons [3 - 5]. The effective potential and vacuum condensate
of four-fermion models on the light-cone have been computed as well [9]. 
Simultaneously, much attention
has been paid for the investigation of dynamical symmetry breaking in 3--dimensional
four-fermion models [5, 6, 10]. There are different reasons for this. 
First, 3--dimensional  theories are directly related to
the high temperature limit of 4--dimensional ones [11, 12]. Second, it has very interesting
non--trivial topological  features [13, 14]. Third, 3--dimensional models are usually 
renormalizable in the large--N  limit [5]. And, finally, there are some hints that they 
could be useful  also in  condensed matter physics [14].

\hspace{-.2cm}
It is quite natural to generalize the known results concerning four-fermion models
for the case of curved spacetime. The phase structure of the NJL model in curved spacetime
turned out to be nontrivial and the phase transition accompanied by the chiral
symmetry breaking has been shown also to take place [15 - 17]. 

The extended NJL model, considered in [4 - 6], generalizes the results for the restricted one
to the case when different kinds of fermion interactions are present in
the effective Lagrangian. This means, strickly speaking, that the Thirring model (TM) [18], 
containing vector--like interactions of fermions, and the NJL model  are combined to
consider their effects simultaneously. 

However the TM is very interesting also by itself. It has been 
investigated in detail from different points of view (see, for example, review [19]).
Its renormalizability has been proved in the $d<4$ dimensions [6]. Its phase structure,
dynamical symmetry breaking and dynamical mass generation have been analysed [20] as well.
Also the gauged TM has been  recently  investigated both in curved [19] and
in flat [19, 21] spacetimes.

In the present paper we are dealing with the three--dimensional massless Thirring model
in curved spacetime. Using the local momentum expansion [22] we obtain the 
linear curvature corrections for the effective potential of the composite auxiliary vector
field. The creation of a non-zero chiral invariant vector-like bifermion 
condensate, taking place  in flat spacetime, is shown to exist
for the model with a finite cutoff above some critical value
of the coupling constant  (a similar phenomenon has been described in [23]). 
The positive  spacetime curvature  induces a second order phase 
transition making the vacuum expectation value of 
$<\overline{\psi}\gamma_\mu\psi>$ tend to zero. The phase structure of the model is discussed. 

\vskip 1cm

{\bf 2}. We start from the action of the three--dimensional Thirring model 
in curved spacetime [19, 20]:
\begin{eqnarray}
S=\int d^3x \sqrt{-g} \left\{i \overline{\psi}\gamma^\mu (x)D_\mu \psi-
\hbox{${\lambda \over 2N}$} (\overline{\psi}\gamma_a \psi)
(\overline{\psi}\gamma^a \psi)\right\},
\end{eqnarray}
where the covariant derivative $D_{\mu}$ is given by
$
D_{\mu}=\partial_{\mu} + \hbox{${1 \over 2 }$} \omega^{ab}{}_{\!\mu} 
\sigma_{ab},
$
the local Dirac matrices $\gamma_\mu (x)$ are expressed through the usual 
flat ones $\gamma_a$ and tetrads $e^a_\mu,\;$
$
\gamma^\mu (x)=\gamma^a e^\mu_a (x), 
\;\;\sigma_{ab}={1\over 4 }[\gamma_a,\gamma_b],
$
the spin--connection has the form 
\begin{eqnarray}
\omega^{ab}{}_{\!\mu}=&\phantom{-}&\hbox{${1 \over 2 }$} e^{a 
\nu}(\partial_{\mu}e^{b}_{\nu}-\partial_{\nu}
e^{b}_{\mu})+\hbox{${1 \over 4 }$} e^{a 
\nu}e^{b\rho}e_{c\mu}(\partial_{\rho}e^{c}_{\nu}
-\partial_{\nu} e^{a}_{\rho})\nonumber\\
&-&\hbox{${1 \over 2 }$} e^{b\nu}(\partial_{\mu}e^{a}_{\nu}-\partial_{\nu} 
e^{a}_{\mu})-
\hbox{${1 \over 4 }$} e^{b\nu}e^{a\rho}e_{c\mu}(\partial_{\rho}e^{c}_{\nu}
-\partial_{\nu} e^{c}_{\rho}),
\nonumber
\end{eqnarray}
and $N$ is the number of bispinor fields $\psi_n$.
\footnote{It should be noted that in the interaction term of (1) we 
don't care for
the difference between Greek and Latin indices, corresponding to the 
curved and 
tangent flat spacetimes. This term can be written using the 
definition 
$A_a=e_a^\mu A_\mu$ in the same form but with the Greek indices 
instead of Latin ones.} 

Introducing the chiral invariant auxiliary field
\begin{eqnarray}
A^a =-\hbox{${\lambda \over N}$}(\overline {\psi} \gamma^a \psi )
\end{eqnarray}
we have the following expression for the action (1):
\begin{eqnarray}
S=\int d^3 x \sqrt{-g} \{ i\overline{\psi}\gamma^\mu D_\mu \psi +
\hbox{${N \over 2\lambda}$}A^a A_a + A_a(\overline{\psi} \gamma^a \psi)\}.
\end{eqnarray}

By a functional integration the effective potential is obtained in  
leading order of large--$N$ expansion according to: 
\begin{eqnarray}
\hbox{${1 \over N }$} \Gamma_{eff}=
-  \hbox{${1 \over 2\lambda }$} \int d^3 x \sqrt{-g}\, {A^a A_a}
-i\ln \det\{ i\gamma^\mu (x)D_\mu+\gamma_a A^a\}.
\end{eqnarray}
Defining now the effective potential for the constant auxiliary field
 $A_a$ as
\begin{eqnarray}
V_{eff} = -\Gamma_{eff}/ \left(N \int d^3 x\sqrt{-g}\right)
\nonumber
\end{eqnarray}
one gets
\begin{eqnarray}
V_{eff}=- \hbox{${1 \over 2\lambda }$} {A_a A^a} + i Sp \ln<x|[ i\gamma^\mu 
(x)D_\mu+\gamma_a A^a]|x>.
\nonumber
\end{eqnarray}

Now we introduce the special  Green function (GF)  obeying the 
following equation: 
\begin{eqnarray}
(i \gamma^\mu D_\mu+\gamma_a A^a -s)_x G(x,x';s)= \delta(x-x').
\end{eqnarray}
Then, neglecting an infinite constant, we find 
\begin{eqnarray}
\ln <x|[i \gamma^\mu D_\mu+\gamma_a A^a]|x>=\int_0^\infty G(x,x;s)ds,
\nonumber
\end{eqnarray}
and for the effective  potential we obtain: 
\begin{eqnarray}
V_{eff}=- \hbox{${1 \over 2\lambda }$} {A_aA^a}+iSp\int_0^\infty G(x,x;s)ds.
\nonumber
\end{eqnarray}

The most convenient way to calculate the linear curvature corrections 
is to use the
local momentum expansion, writing the spacetime quantities contained in 
the equation (5)  by the help of Riemannian normal coordinates   
up to linear curvature accuracy in the following form [22]: 
\begin{eqnarray}
g_{\mu\nu}(x)&=&\eta_{\mu\nu}-\hbox{${1 \over 3 }$} R_{\mu\rho\sigma\nu}y^\rho 
y^\sigma \qquad \hbox{with}\qquad  y=x-x',\nonumber\\
e^\mu_ a (x)&=&\delta^\mu_a+\hbox{${1 \over 6 }$} R^\mu{}_{\!\rho\sigma a}y^\rho 
y^\sigma, \qquad
\omega^{ab}{}_{\!\mu}\sigma_{ab}
=\hbox{${1 \over 2 }$} R^{ab}{}_{\!\mu\lambda} 
y^\lambda\sigma_{ab}.
\nonumber
\end{eqnarray}

Substituting these expressions into (5) we obtain the equation for the 
Green function $G(x,x';s)$:
\begin{eqnarray}
\biggl[ i\gamma^a (\delta^\mu_a+\hbox{${1 \over 6 }$} R^\mu{}_{\!\rho\sigma 
a}y^\rho y^\sigma)
(\partial_\mu&+&\hbox{${1 \over 4 }$}
R_{bc\mu\lambda}y^\lambda\sigma^{bc})\nonumber\\
&+&A_a\gamma^a-s \biggr] 
G(x,x',s)=\delta(x-x')
\nonumber
\end{eqnarray}

Performing the expansion according to the degree of the spacetime 
curvature, 
$
G=G_0+G_1+. . . ,
$
where $G_n \sim R^n$, we receive the following chain of equations:
\begin{eqnarray}
(i\not{\!\partial}+\not{\!\!A}-s) G_0(x,x',s)&=&\delta(x-x'),\\
(i\not{\!\partial}+ \not{\!\!A}-s)G_1(x,x',s)&+&\nonumber\\
+i\,\biggl[\hbox{${1\over 6 }$}
\; R^\mu{}_{\!\rho\sigma a}y^\rho y^\sigma \gamma^a\partial_\mu &+&
\hbox{${1 \over 4 }$} R_{bca\lambda} y^\lambda\gamma^a \sigma^{bc} 
\biggr]G_0(x,x',s)=0.\hspace{2cm}
\end{eqnarray}
Here and below we can forget about the difference between Greek and Latin
indices because it lies beyond of linear curvature approximation.

Equations (6) and (7) can be solved directly in the momentum 
representation:
\begin{eqnarray}
G_0(k)=& &\frac{\not{\!q}+s}{q^2-s^2},\\
G_1(k)=&-&\frac{1}{12}\frac{\not{\!q}+s}{(q^2-s^2)^2} + 
\frac{2}{3}R^{\mu\nu}\frac{k_\mu k_\nu(\not{\!q}+s)}{(q^2-s^2)^3}-
\frac{4}{3}R^{\mu\rho\sigma\nu}\frac{k_\mu\gamma_\nu A_\rho 
q_\sigma}{(q^2-s^2)^3}
\nonumber\\
&-&\frac{1}{2}R^{\mu\nu\rho\lambda}\frac{\gamma_\rho\sigma_{\mu\nu}q_\lambda}{(q^2
-s^2)^2}-
\frac{1}{3}R^{\mu\rho\sigma\nu}\frac{(\not{\!q}+s)k_\mu\gamma_\nu(\gamma_\rho
A_\sigma+
\gamma_\sigma A_\rho)}{(q^2-s^2)^3}\nonumber\\
 &+&\frac{1}{3}R^{\mu\nu}\frac{\gamma_\mu A_\nu}{(q^2-s^2)^2}+
\frac{1}{3}R^{\mu\nu}\frac{A_\mu(A_\nu+4k_\nu)(\not{\!q}+s)}{(q^2-s^2)^3},
\end{eqnarray}
where
$
q_\mu=k_\mu+A_\mu.
$

Let us now, for simplicity, assume that our spacetime has constant 
curvature such that
\begin{eqnarray}
 R_{\mu\sigma\kappa\lambda}=\hbox{${1 \over 6 }$} R \biggl(\eta_{\mu\kappa}
 \eta_{\sigma\lambda}-\eta_{\mu\lambda}\eta_{\kappa\sigma} \biggr), \qquad
R_{\mu\nu}=\hbox{${1 \over 3 }$}R \eta_{\mu\nu}.
\nonumber
\end{eqnarray}

Then, calculating the trace over the spinor indices (the dimension of the
fermion representation is supposed to  be equal to four)  and making a Wick 
rotation, one gets:
\begin{eqnarray}
V_{eff}= \frac{A^2}{2\lambda}&-& 2\int_0^\infty ds^2\int\frac{d^3 
k}{(2\pi)^3} 
\Biggl[ -\frac{1}{(k+A)^2+s^2}+\nonumber\\
&+&\frac{R}{36}\left(\frac{3}{((k+A)^2+s^2)^2}
+2\frac{k^2-A^2-3s^2}{((k+A)^2+s^2)^3}\right) \Biggr],
\nonumber
\end{eqnarray}
where $A^2=A_a A^a$ is the Euclidean square. Integrating over
 the angles (in momentum space), we finally get the expression 
for the effective potential in the model under consideration:
\begin{eqnarray}
V_{eff}={\frac{A^2}{2\lambda}}
&-&{\frac{1}{2\pi^2}}\Bigg\{{\frac{3\Lambda^4+6A^2\Lambda^2-A^4}{12A}}
\ln\left|{\frac{\Lambda+A}{\Lambda-A}}\right|\\
&+&
{\frac{2}{3}}\Lambda^3\ln\left|{\frac{\Lambda^2-A^2}{\Lambda^2}}\right|
+{\frac{1}{6}}\Lambda A^2-
{\frac{R}{36}}A\ln\left|{\frac{\Lambda+A}{\Lambda-A}}\right| \Bigg\},
\nonumber
\end{eqnarray}
where the parameter $\Lambda$ is  the upper  cut off in the above momentum 
integral. 

There are two ways of explaining the introduction of this parameter. The 
first one is the standard renormalization  cut off. Then, after 
renormalization of the coupling constant $\lambda$ 
in the limit $\Lambda\to\infty$,
\begin{eqnarray}
\frac{1}{\lambda_R}=\frac{1}{\lambda}-\frac{2\Lambda}{3\pi^2},
\end{eqnarray}
we have
\begin{eqnarray}
V_{eff}=\frac{A^2}{2\lambda_R}
\end{eqnarray}
and, hence, the Thirring model is the trivial one in this 
approximation.

However, in general,  four-fermion models may be considered  as
 low energy effective theories being derived from a more
 complete version of quantum field theory (QCD, for
example). In this case the parameter $\Lambda$ can be treated as a 
natural characteristic
scale limiting the range, where our low energy approximation is valid,
and then the Thirring model would 
describe some phenomenological effects of elementary particles physics. 

Keeping in mind these reasoning we can save $\Lambda$ finite and 
obtain:
\begin{eqnarray}
V''_{eff}(0)=\frac{1}{\lambda}
-\frac{2\Lambda}{3\pi^2}+\frac{R}{18\Lambda\pi^2},
\qquad
V'''_{eff}(0)=0.
\end{eqnarray}
The  creation of a non-zero vacuum expectation value of the
vector--like bi\-fermio\-nic condensate
$
A^a\sim <(\overline{\psi}\gamma^a\psi)>
$
takes place if
$
V''_{eff}(0)<0.
$

For $R=0$  the critical value of coupling constant $\lambda$ is given by
\begin{eqnarray}
\lambda_c=\frac {3\pi^2}{2\Lambda}.
\end{eqnarray}
 Therefore, if $\lambda>\lambda_c$, a nonvanishing condensate $A_a\ne 0$ appears.
However, positive $R$ makes $A_a$ tend to zero if 
\begin{eqnarray}
R>8\Lambda^2\left( 1- \frac{\lambda_c}{\lambda}\right).
\end{eqnarray}
Furthermore, the equation of the critical line dividing the 
$R - \lambda$ plane  into two regions, 
characterized by 
zero and non-zero vacuum expectation value of
 $<\overline{\psi}\gamma_a\psi>$, is determined by:
\begin{eqnarray}
\frac{R}{8\Lambda^2}+\frac{\lambda_c}{\lambda}=1.
\end{eqnarray}
Because of condition (15) we have the ordinary picture of second order 
curva\-ture-induced 
phase transition where  $A^2$ is the  order paremeter.
It is clearly illustrated by  Figs. 1 - 3.

Fig. 1 shows the dependence of $v(a)=[V(a)- V(0)]/{\Lambda^3}$
on $a=A/\Lambda$  for $R= 0$. The curves 1, 2, 3, 4 correspond to the
 following 
values of 
$g=\lambda\Lambda=12.5, 15.25, 16.5, 17$.

Fig. 2 represents the behaviour of the function
 $v(a)$  (curves 1, 2, 3, 4) for different 
values of the curvature $r=R/\Lambda^2=0.8, 0.5, 0.3, 0$ for $g= 17$.

Fig. 3 contains the three--dimensional plot of effective potential
 $v(a,r)$ as a function of
$a$ and $r$ simultaneously for $g= 17$ as well.
\vskip 1cm

{\bf 3.} The authors would like to thank S. D. Odintsov for pointing  out  the
problem and for helpful discussions.
Yu. I. Sh. is very much indebted to Deutcher Akademischer Austauschdienst 
for financial support of his  visit to Leipzig 
University, where  this work has been done. He  also expresses his 
deep gratitude to 
A. Letwin and R. Patov for their  kind support.

\vfill\eject
\begin{figure}[h]
\centerline{\epsfig{figure=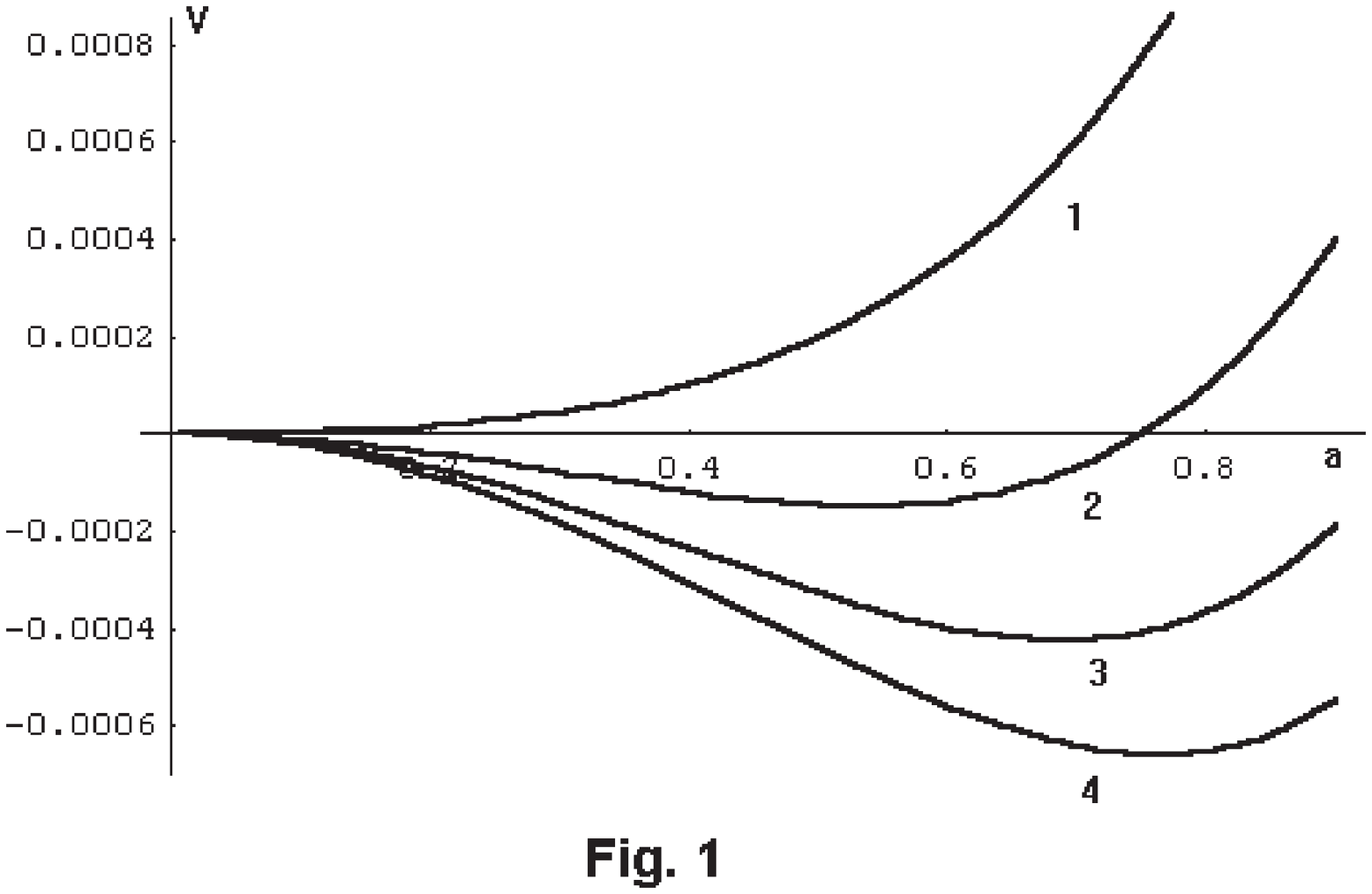, scale=.5}}
\end{figure}
\pagebreak
\begin{figure}[h]
\centerline{\epsfig{figure=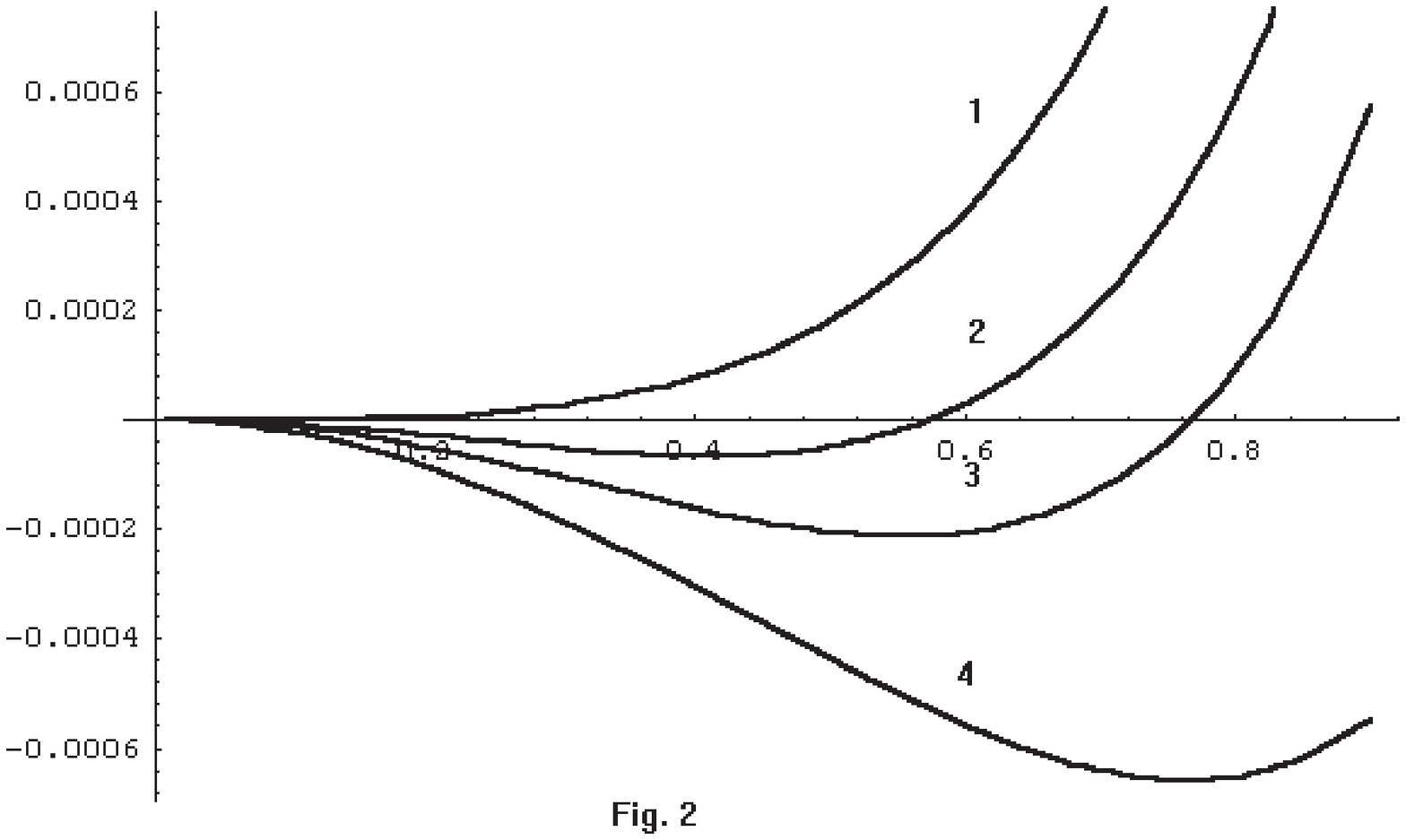, scale=.5}}
\end{figure}
\pagebreak
\begin{figure}[h]
\centerline{\epsfig{figure=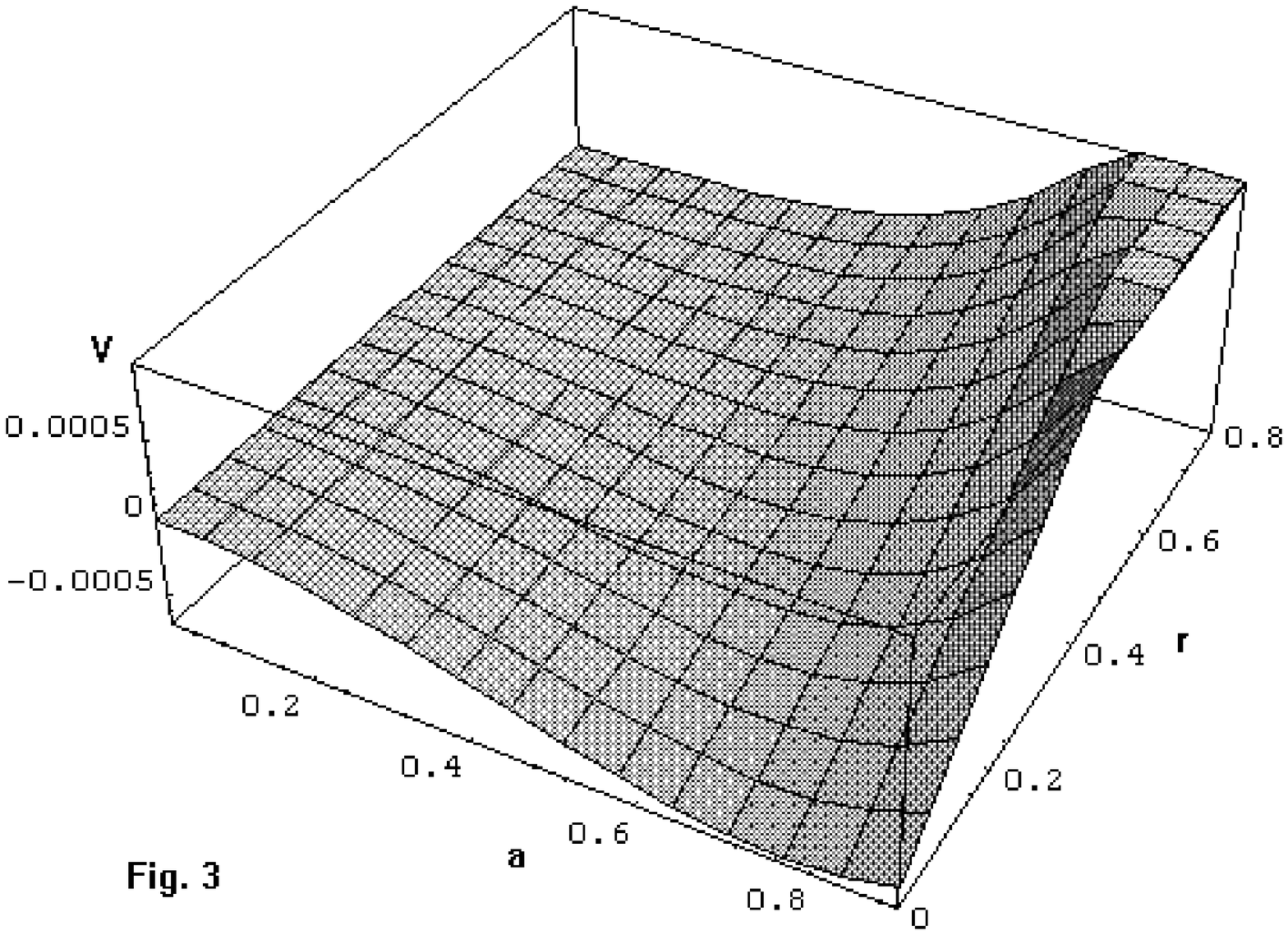, scale=.5}}
\end{figure}

\end{document}